# Energy gap and London penetration depth of $MgB_2$ films determined by microwave resonator measurements

N. Klein, B.B. Jin, J. Schubert, M. Schuster, H.R. Yi

Forschungszentrum Jülich, Institute of Thin Films and Interfaces, D-52425 Jülich, Germany,

A. Pimenov, A. Loidl

Universität Augsburg, Experimentalphysik V, EKM, 86135 Augsburg, Germany

S.I. Krasnosvobodtsev

P.N. Lebedev Physics Institute, Russian Academy of Sciences, 117924 Moscow, Russia

*Abstract* - We have measured the temperature dependence of the microwave surface impedance $Z_s = R_s + i\omega\mu_0\lambda$ of a $MgB_2$ film at a frequency $\omega/2\pi$ of 18 GHz employing a dielectric resonator technique. We found that the temperature dependence of the magnetic field penetration depth $\lambda$ can be fitted by $\lambda(T) = \lambda(0) [1-(T/T_c)^2]^{-1/2}$ with $\lambda(0) = (260 \pm 20)$ nm. The absolute value of $\lambda(0)$ was confirmed by direct measurements employing submillimeter wave transmission spectroscopy at 430 GHz. The analysis of the $\lambda(T)$ data below $T_c/2$ revealed significant deviations from the quadratic temperature dependence. In contrast, we found that an exponential temperature dependence fits the experimental data within the statistical measurement error for temperature changes of $\lambda$ of $\pm$ 0.4 nm. This observation indicates thermal excitation of quasiparticles over a finite energy gap of $(3.3 \pm 0.3)$ meV corresponding to $\Delta/kT_c = 1.2 \pm 0.1$. Our results strongly supports the existence of a multiple gap or a strongly anisotropic gap and the absence of nodes in the gap function.

The question about the energy gap in a particular superconducting material is of fundamental importance for the understanding of the relevant pairing mechanism and for the determination of its application potential. The recently discovered superconductivity in $MgB_2$ [1] raises this question with particular emphasis. According to initial findings, $MgB_2$ seemed to comprise a high transition temperature with superconducting properties resembling that of conventional superconductors rather than that of high temperature superconducting cuprates [2]. In particular, the strong anisotropy of superconducting properties, the short coherence length and the unconventional order parameter being apparent for high temperature superconducting cuprates have turned out to be a burden for many applications. Consequently, such burdens are expected not to be present or at least less pronounced in $MgB_2$.

Recent experiments have brought some clarification about the gap features of $MgB_2$. According to tunneling spectroscopy [3], point contact spectroscopy [4] and Raman scattering [5] there is evidence for an s-wave symmetry with two distinct gaps possibly associated with the two separate segments of the Fermi surface being present in $MgB_2$ [6]. The width values of these two gaps were determined to be around 2.0 to 3.0 meV and 6.2 to 7.5 meV consistently by these different methods.

On the other hand, the reported quadratic temperature dependence of the penetration depth $\lambda(T)$ indicates unconventional superconductivity similar to high temperature superconducting cuprates [7,8]. Such finding would imply either nodes in the gap function or a finite density of quasiparticle states at the Fermi energy due to the existence of a non superconducting subband [9]. However, for the case of a very small minimum gap $\Delta_{min}$ as suggested by the above mentioned experimental findings a very high measurement accuracy for temperature changes of $\lambda$ at $T<<T_c$ is required to distinguish between a quadratic and a $\exp(-\Delta_{min}/kT)$ – like temperature dependence.

Microwave surface impedance measurements have proved to be the most sensitive tool to determine the temperature dependence of the magnetic field penetration depth of both thin film [10] and bulk single crystal samples [11]. In particular, they have been employed successfully to attain significant information about the symmetry of the order parameter in the high temperature superconducting cuprates [10-12]. Therefore, microwave resonator techniques are most appropriate to be used for high-precision $\lambda(T)$ measurements on $MgB_2$ samples.

The $MgB_2$ films were grown by two-beam laser ablation on a plane parallel [1102] oriented sapphire substrate of 10x10mm$^2$ in size [13]. The films exhibit sharp superconducting transitions at $T_c$ =32 K. The thickness of the film was determined to be (600 ± 50) nm by use of an $\alpha$ - stepper at one of the small holes (diameter around 30 µm) being apparent in the films.

The microwave surface impedance was determined using a sapphire dielectric resonator technique described elsewhere [10,14]. The cavity with part of one endplate replaced by the thin film sample was excited in the $TE_{01\delta}$ - mode under weak coupling conditions. The unloaded quality factor $Q_0$ and resonant frequency was recorded as a function of temperature. The real part of the surface impedance, the surface resistance $R_s$ was determined according to

$$R_s(T) = G\left[\frac{1}{Q_0(T)} - \frac{1}{Q_{niobium}(4.2K)}\right] \qquad \text{Eq. 1}$$

with $G = 740\ \Omega$ being a geometrical factor determined by numerical simulation of the electromagnetic field distribution in the cavity and $Q_{niobium}(4.2\ K) = 92000$ representing the unloaded qualtity factor measured by employing a high-quality niobium thin film as sample. For temperature below 30 K Eq. 1 allows for the determination of $R_s$ with a systematic error of about 0.1 m$\Omega$, which is due to neglecting the temperature dependent background losses of the cavity and the small microwave losses ($R_s \approx 10^{-5}\Omega$) of the niobium film.

The temperature dependence of the penetration depth was determined from the temperature dependence of the resonant frequency $f(T)$ using

$$\delta\lambda(T) = \frac{G}{\pi\mu_0}\frac{f(T) - f(4.2K)}{f^2(4.2K)} \qquad \text{Eq. 2}$$

with $\mu_0 = 1.256 \cdot 10^{-6}$ Vs/Am. According to Eq. 2 a change of $\lambda$ by 1 nm corresponds to a frequency change of 1.7 kHz, which is a fraction of $6 \cdot 10^{-3}$ of the halfwidth of the resonance curve ($Q_0 \approx 70000$). We have found that our statistical measurement error is about 0.6 kHz, if several frequency sweeps are averaged. Therefore, our resolution for $\delta\lambda$ is about 0.4 nm. Similar to $R_s$ there is a systematic error due to frequency changes caused by the thermal expansion and the temperature dependence of the skin depth of the cavity wall material (copper). To account for this effects, we recorded $f(T)$ employing a copper sample. From this measurement we found that the systematic error is less than 2.5 nm for $T \leq 30K$ and less than 1 nm for $T \leq 15$ K. Therefore, Eq. 2 was applied without correction for our investigation.

In general, microwave resonator measurements do not allow for the determination of absolute values of $\lambda$, because the resonator dimensions are only know with a precision of several ten micrometers. In order to confirm the $\lambda(0)$ values as determined from a fit of a theoretical temperature dependence, we employed a quasioptical submillimeter wave transmission spectroscopy at $f = 430$ GHz as described in [8]. This method allows an absolute determination of the complex conductivity $\sigma = \sigma_1 + i\sigma_2$ in the superconducting state, provided that the film thickness is known. The absolute value of $\lambda(0)$ was determined by $\sigma_2(T \to 0)$

=$[\omega\mu_0\lambda(0)^2]^{-1}$ with an accuracy determined by the accuracy of film thickness determination or possible thickness inhomogeneity. In addition, the measured values of $1/\sigma_1$ just above $T_c$ is of the order of the normal state dc resistivity, which is an important figure-of-merit for the quality of the films.

Fig. 1 shows surface resistance and the penetration depth as determined by Eq. 1 and Eq. 2, respectively. The flat temperature dependence of $R_s$ at $T \ll T_c$ indicates the existence of normal conducting phases in the film leading to a finite residual surface resistance, very similar to the one observed for high-temperature oxide superconducting films [15]. Such a high residual resistance is supposed to hide a possible exponential temperature dependence due to a finite energy gap, as observed for niobium. The inset of Fig. 1 represents the complex submillimeter wave conductivity. The value of $1/\sigma_1$ at $T = 40$ K corresponds to a dc resistivity of 49 µΩcm, which is a quite low value for $MgB_2$ films. From the value of $\sigma_2$ at $T = 4.2$ K we determined the absolute value of $\lambda_0$ to be $(290 \pm 30)$ nm.

In contrast to the surface resistance, the penetration depth is much less sensitive to normal conducting impurities, because it probes the density of Cooper pairs rather than the density of normal conducting charge carriers in $R_s$, which are composed out of thermally excited quasiparticles plus impurity related normal conducting charge carriers. In addition, as indicated by the scattering of the data point, temperature changes of $\lambda$ can be determined with a much higher precision than of $R_s$.

Fig.2 shows the measured temperature dependence of $\lambda$ determined from $f(T)$ according to Eq. 2. Apart from small deviations close to $T_c$ the data can be fitted by

$$\delta\lambda(T) = \lambda_0 f(T) \coth\left[\frac{t}{\lambda_0 f(T)}\right] - \lambda_0 \coth\left[\frac{t}{\lambda_0}\right] \qquad \text{Eq. 3}$$

with $t$ being the thickness of the film. The "coth" - term represents a correction to account for the fact that the penetration depth is of the order of the film thickness [16]. The function $f(T)$ contains all the information on the thermal excitation spectrum of the quasiparticles. According to BCS theory in the weak coupling limit $f(T)$ is given by

$$f(T) = \left\{1 - \left[2\int_{\Delta(T)}^{\infty} -\frac{\partial F(\varepsilon)}{\partial \varepsilon} \frac{\varepsilon}{\sqrt{\varepsilon^2 - \Delta^2(T)}}\right]^2\right\}^{-1/2} \qquad \text{Eq. 4}$$

with $F(\varepsilon)=[\exp(\varepsilon/kT)+1]^{-1}$ representing the Fermi function and $\Delta(T)$ the temperature dependent energy gap according to weak coupling BCS theory [17]. For $T<<T_c$ Eq. 4 can be approximated by

$$f(T) \approx \sqrt{\frac{\pi}{2} \frac{\Delta}{kT_c} \frac{T_c}{T}} \exp\left[-\frac{\Delta}{kT_c} \frac{T_c}{T}\right] \qquad \text{Eq.5}$$

with $\Delta$ representing the energy gap value for $T\rightarrow 0$. For strong coupling superconductors Eq. 5 is still valid, but at higher temperatures significant deviation from Eq. 4 occur. For both cases, the analytical approximation

$$f(T) = \left\{1-\left[\frac{T}{T_c}\right]^n\right\}^{-1/2} \qquad \text{Eq.6}$$

with $n(T) = 3 - T / T_c$ for weak coupling (dashed lines in Figs. 2 and 3) and $n = 4$ for strong coupling (dashed-dotted lines in Fig. 2 and 3) provides a reasonable fit to numerically calculated data at higher temperatures (typically about $T_c/2$) [18]. According to experimental results on high temperature superconducting cuprates and calculations in the framework of a $d$-wave pairing state $n = 2$ represents a reasonable approximation for an unconventional pairing state in the case of a high level of impurities (dotted lines in Fig. 2 and 3) [9]. In any case, the occurrence of an exponential behaviour of $\delta\lambda(T)$ at $T << T_c/2$ according to Eq. 5 is a clear indication for a finite energy gap in all directions of the Fermi surface without nodes (full lines in Figs. 2 and 3). In the possible case of an anisotropic or multicomponent gap the exponential temperature dependence expressed by Eq. 5 still remains valid if $\Delta$ is replaced by the minimum value of the gap.

According to Fig. 2 $n = 2$ gives the best fit to our data for $\lambda_0 = (260 \pm 20)$ nm. This is in good agreement with the absolute value of $(290 \pm 30)$ nm determined by submillimeter wave spectroscopy. It should be noted that this value is more than a factor of two higher then the one determined by muon spin rotation on bulk samples [7]. The small deviations of $\delta\lambda(T)$ from the fit close to $T_c$ result from the additional screening of the quasiparticles at microwave frequencies, i.e. from a smooth transition of the London penetration depth below $T_c$ to the normal metal skin depth above $T_c$.

Fig. 3 shows a magnification of the $\delta\lambda(T)$ data for $T \leq 15$ K. Within the measurement accuracy, significant deviations from the quadratic temperature dependence occur. In contrast, an exponential temperature dependence according to Eq. 5 gives the best fit for $T \leq 16$ K assuming $\Delta/kT_c = 1.2 \pm 0.1$ corresponding to $\Delta = (3.3 \pm 0.3)$ meV. Fig. 4 shows the low temperature part of the surface resistance and a fit of

$$R_s(T) = A \exp\left[\frac{-\Delta}{kT}\right] + R_{res} \qquad \text{Eq. 7}$$

corresponding to a superconductor with a finite gap and metallic impurities leading to a temperature independent residual surface resistance $R_{res}$. In contrast to the fit of $\delta\lambda(T)$ the effect of the finite film thickness can be neglected, because its effect on the temperature dependence is negligible within the measurement accuracy of $R_s(T)$. According to Fig. 4 Eq. 7 fits the experimental data employing $\Delta = 3.3$ meV as determined from $\delta\lambda(T)$. It should be emphasised that the accuracy of $R_s(T)$ is too low to distinguish between a quadratic and an exponential dependence.

In conclusion, our experimental finding represents a clear evidence for the existence of a finite gap in all directions of the Fermi surface. The small value of $\Delta/kT_c = 1.2$ indicates that this gap value represents a minimum value of a strongly anisotropic or multicomponent gap. Our results definitely opposes the occurrence of nodes in the gap. We speculate, that the agreement with a quadratic temperature dependence ($n = 2$ in Eq. 6) above $T_c / 2$ indicates a gradual increase of the quasiparticle density-of-states for energies above the observed $\Delta_{min}$ of 3.3 meV up to a peak at a possible maximum value of the energy gap.

**Figure captions:**

Fig. 1: Temperature dependence of the surface resistance $R_s$ and temperature changes of the magnetic field penetration depth $\delta\lambda$ of a MgB$_2$ thin film recorded at 18 GHz using a dielectric resonator technique. The inset shows the complex conductivity at $f = 430$ GHz determined by submillimeter wave spectroscopy.

Fig. 2: $\delta\lambda(T)$ data from Fig. 1 and results of fits of theoretical dependencies as described in the text.

Fig. 3: Magnified view of Fig. 2 for the temperature range below 22 K.

Fig. 4: Low temperature region of $R_s(T)$ with an exponential fit as described in the text using the value of $\Delta/kT_c$ derived from the fit to $\delta\lambda(T)$.

Fig. 1 (top ) and Fig. 2 (bottom)

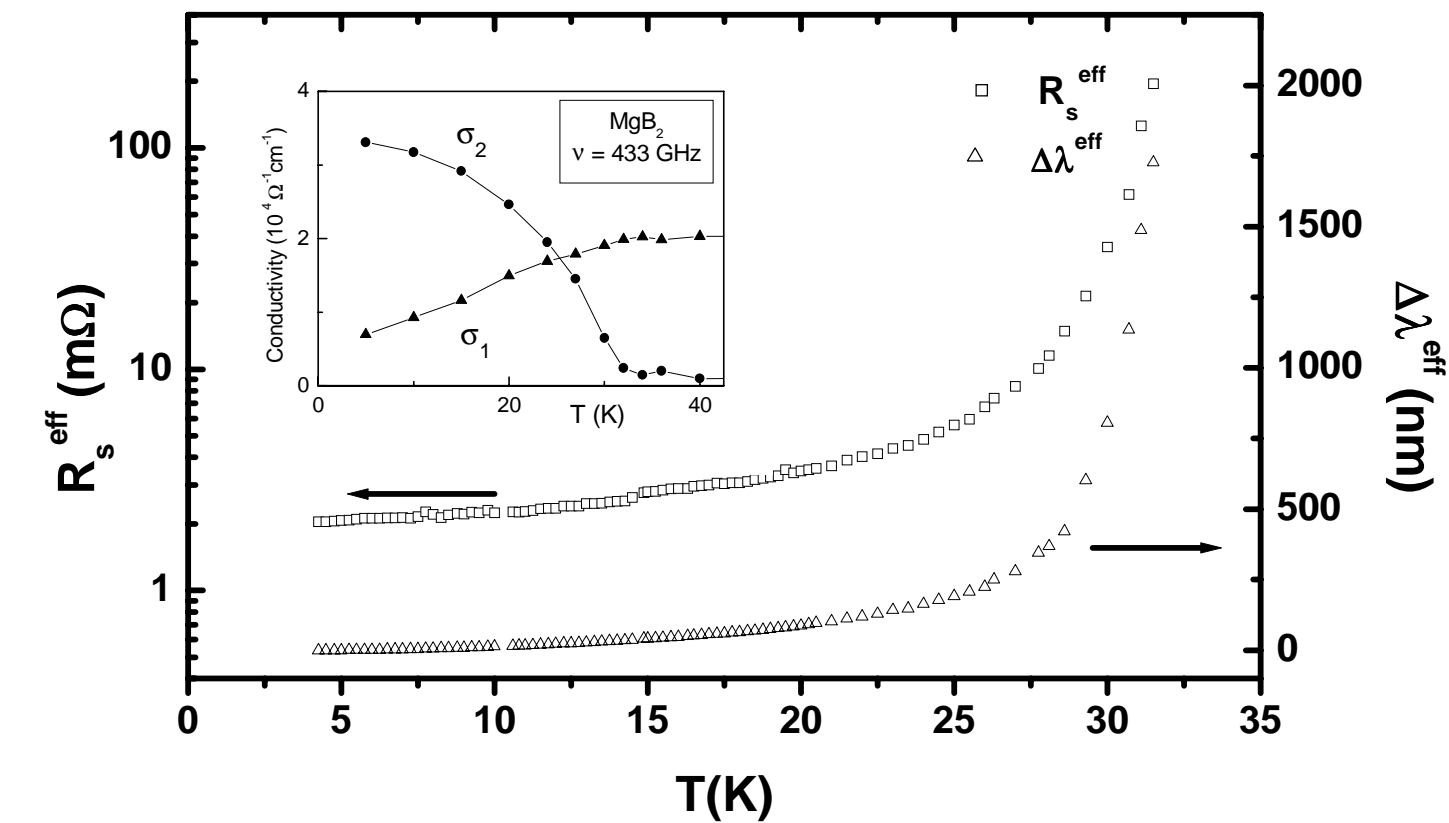

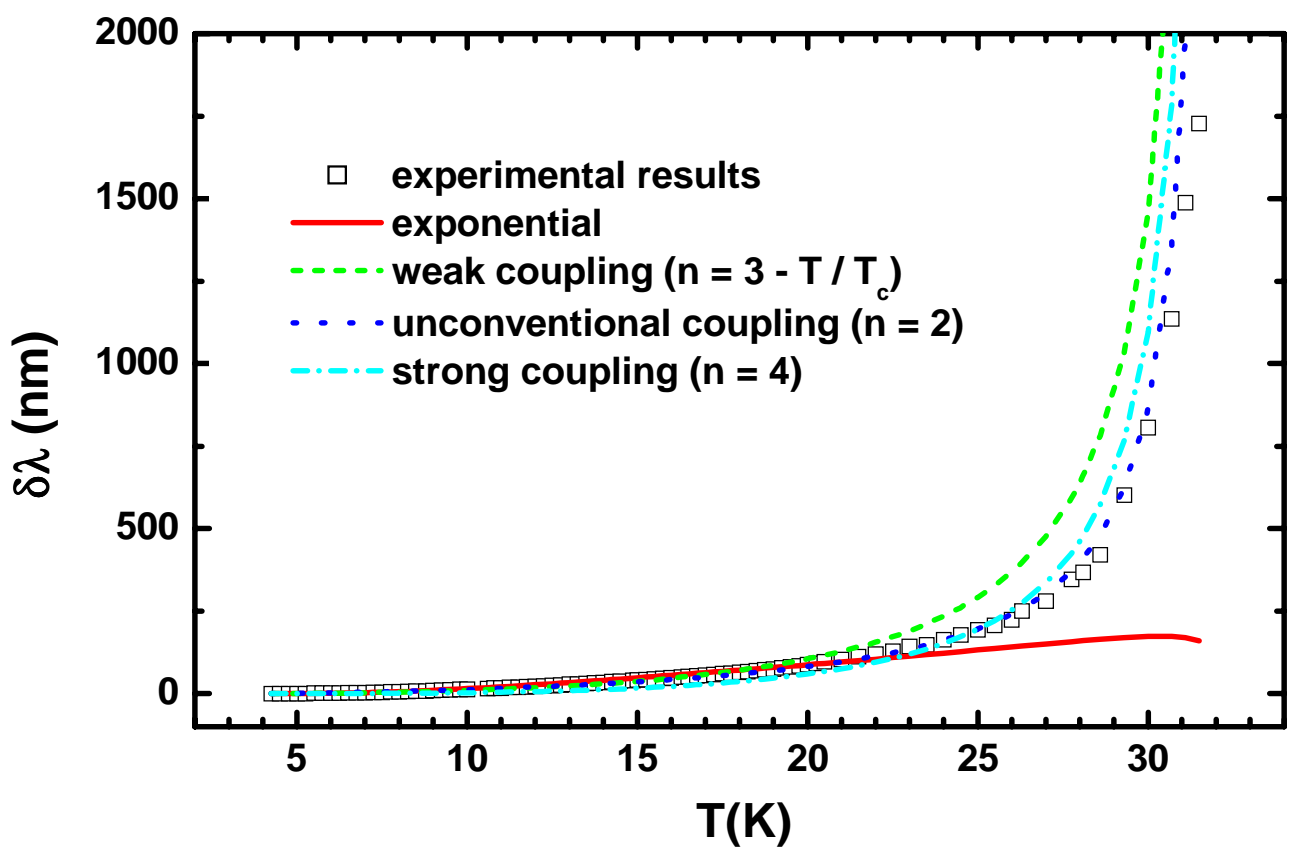

Fig. 3 (top ) and Fig. 4 (bottom)

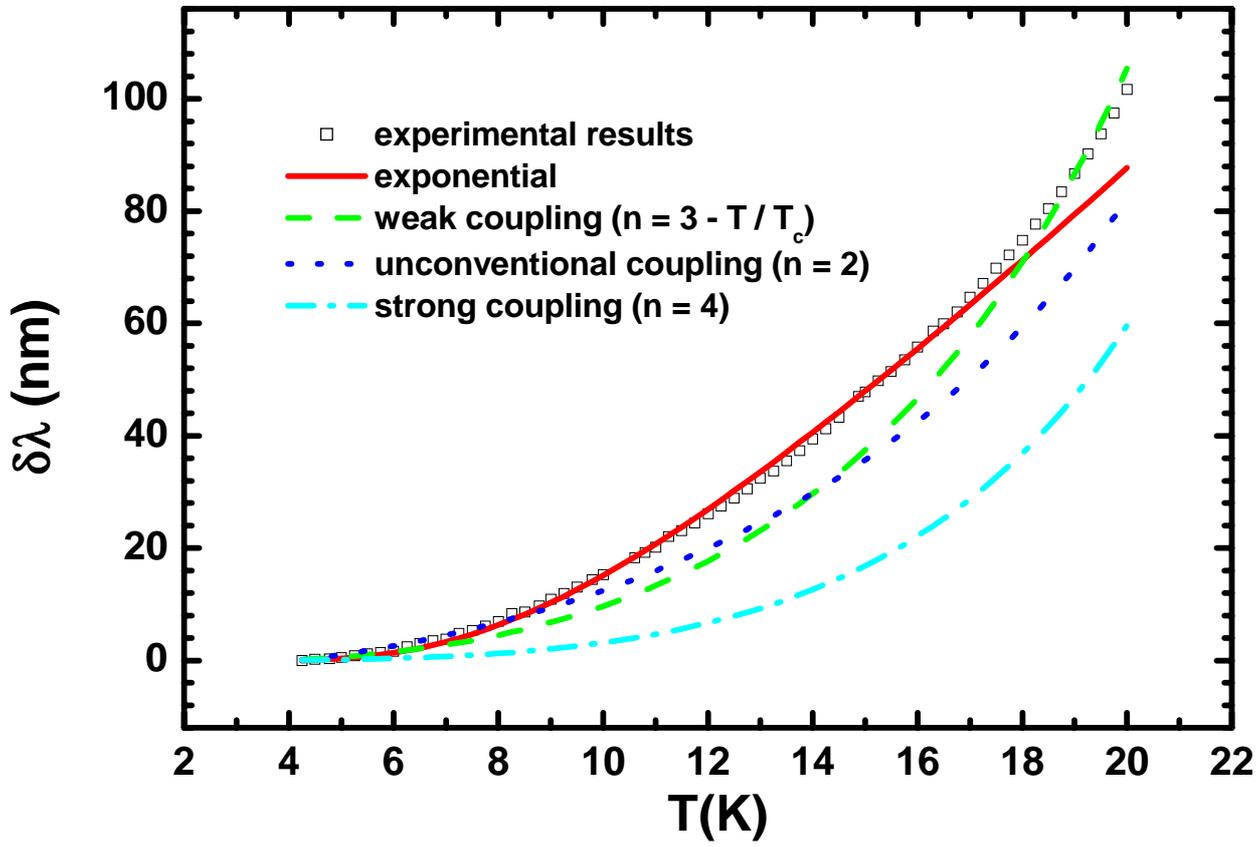

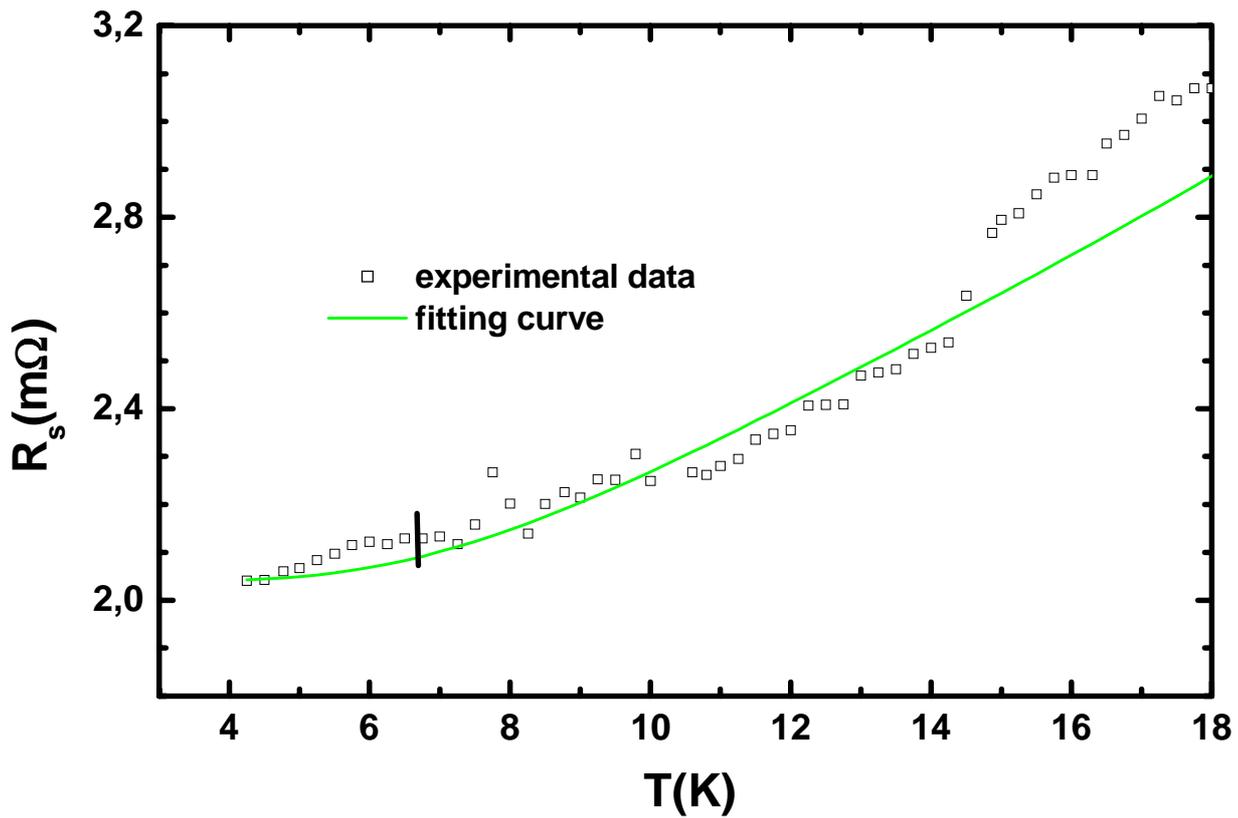